\documentclass[lettersize,journal]{IEEEtran}
\usepackage{amsmath,amsfonts}
\usepackage{algorithmic}
\usepackage{algorithm}
\usepackage{array}
\usepackage[caption=false,font=normalsize,labelfont=sf,textfont=sf]{subfig}
\usepackage{textcomp,amssymb}
\usepackage{stfloats}
\usepackage{url}
\usepackage{verbatim}
\usepackage{graphicx}
\usepackage{cite}
\begin{document}

\title{A Novel HARQ-CC Assisted SCMA Scheme}

\author{Man Wang, Zheng Shi, Yunfei Li, Xianda Wu, Weiqiang Tan, and Xinrong Ye
\thanks{The corresponding author is Zheng Shi.}
\thanks{Man Wang and Zheng Shi are with the School of Intelligent Systems Science and Engineering, Jinan University, Zhuhai 519070, China (e-mails: wang0906@stu2022.jnu.edu.cn, zhengshi@jnu.edu.cn).}
\thanks{Yunfei Li is with the Department of Electrical Engineering, Anhui Polytechnic University, Wuhu City, China (email: lyf@mail.ahpu.edu.cn).}
\thanks{Xianda Wu is with the School of Electronics and Information Engineering, South China Normal University, Foshan 528000, China (email: xiandawu@m.scnu.edu.cn).}
\thanks{Weiqiang~Tan is with the School of Computer Science and Cyber Engineering, Guangzhou University, Guangzhou 510006, China  (e-mail: wqtan@gzhu.edu.cn).}
\thanks{Xinrong~Ye is with the School of Physics and Electronic Information, Anhui Normal University, Wuhu 241002, China (e-mail: shuchong@mail.ahnu.edu.cn).}
}

\markboth{Journal of \LaTeX\ Class Files,~Vol.~14, No.~8, August~2021}%
{Shell \MakeLowercase{\textit{et al.}}: A Sample Article Using IEEEtran.cls for IEEE Journals}


\maketitle

\begin{abstract}

This letter proposes a novel hybrid automatic repeat request with chase combining assisted sparse code multiple access (HARQ-CC-SCMA) scheme. Depending on whether the same superimposed packet are retransmitted, synchronous and asynchronous modes are considered for retransmissions. Moreover, factor graph aggregation (FGA) and Log-likelihood ratio combination (LLRC) are proposed for multi-user detection. Regarding FGA, a large-scale factor graph is constructed by combining all the received superimposed signals and message passing algorithm (MAP) is applied to calculate log-likelihood ratio (LLR). Whereas, owing to the same unsuccessful messages required to be retransmitted, LLRC adds up LLRs of erroneously received packets in previous HARQ rounds together with currently received packets for joint channel decoding and LLRs of failed users are saved. Finally, Monte Carlo simulations are preformed to show that FGA surpasses LLRC and HARQ with incremental redundancy (HARQ-IR) in synchronous mode. However, LLRC performs better than FGA at low signal-to-noise ratio (SNR) in asynchronous mode, because failed messages after the maximum allowable HARQ rounds in this mode yields significant error propagation in low SNR regime.

\end{abstract}


\section{Introduction}
\IEEEPARstart{T}{he} sixth generation (6G) of wireless networks faces challenges of ultra-high data rate, low latency, high reliability, and massive connections for emerging applications of metaverse, digital twins, sensory interconnection, holographic communication, etc. \cite{10041914}. Non-orthogonal multiple access (NOMA) is one of the key enablers to accommodate these demands. NOMA can be further divided into two categories based on the type of shared resources, i.e., power-domain NOMA and code-domain NOMA\cite{9409837}. As a representative class of code-domain NOMA, sparse code multiple access (SCMA) has been recognized as a promising multiuser multiplexing technique, because multi-dimensional codewords brings constellation shaping gain and the sparsity yields a reduced decoding complexity\cite{9782313}. 

However, SCMA cannot guarantee the reception reliability while suffering from bad propagation conditions. To address this issue, hybrid automatic repeat request (HARQ) can be invoked to improve the reliability of SCMA \cite{7500115,8322726,10295166,zhu2018rateless,e25060930}. The essence of HARQ is to retransmit erroneously received packets, which results in an enhanced reliability. HARQ can be classified into Type-I HARQ, HARQ with chase combining (HARQ-CC), and HARQ with incremental redundant (HARQ-IR) according to different coding and combining schemes\cite{9745558}. Recently, a few HARQ assisted SCMA (HARQ-SCMA) schemes have been proposed in the literature. For example, Type-I HARQ aided SCMA was studied in \cite{7500115}, where some successfully received messages are retransmitted to facilitate codebook design as well as reduce decoding complexity. Moreover, a codebook hopping scheme and a network-coding aided K-repetition scheme were designed for HARQ-SCMA in \cite{8322726} and \cite{10295166}, respectively. 
Furthremore, HARQ-IR assisted SCMA (HARQ-IR-SCMA) was examined in \cite{zhu2018rateless}, where multi-user detection and channel decoding are iteratively repeated till convergence. 
In order to further improve the reliability, a windowed joint detection and decoding algorithm was proposed for HARQ-IR-SCMA in \cite{e25060930}, wherein extrinsic information between the channel decoder and $w$ multi-user detectors are exchanged. 
It is noteworthy that HARQ-IR performs the best among the three HARQ schemes, albeit at the cost of highest incremental decoding complexity. In HARQ-IR-SCMA systems, the iterations between incremental decoding and multi-user detection result in prohibitively high computational burden. Accordingly, we focus on HARQ-CC assisted SCMA (HARQ-CC-SCMA), which is expected to achieve a balanced tradeoff between complexity and reliability.

In this letter, we propose a novel HARQ-CC-SCMA scheme. On the basis of whether the same superimposed packet of all the users are resent, synchronous and asynchronous retransmission modes are considered. Moreover, two multi-user detection schemes are developed to combine the current packet with the erroneous ones, i.e., factor graph aggregation (FGA) and Log-likelihood ratio combination (LLRC). More specifically, with regard to FGA, a large-scale factor graph is constructed by using all the received superimposed signals. Message passing algorithm (MPA) is then leveraged to compute LLRs of different users. Whereas, owing to the same unsuccessful messages delivered in retransmission, LLRC adds up Log-likelihood ratios (LLRs) of failed message received in all HARQ rounds for channel decoding and stores the LLRs of erroneous messages after decoding. In the end, simulation results are exhibited to reveal that FGA outperforms LLRC in synchronous mode. Furthermore, FGA is superior to HARQ-IR-SCMA in synchronous mode. Under such a circumstance, joint SCMA detection among HARQ rounds before channel decoding is more powerful than the joint channel decoding among HARQ rounds after SCMA. Besides, LLRC performs better than FGA at low signal-to-noise ratio (SNR) in asynchronous mode. This is because failed messages after the maximum allowable HARQ rounds in asynchronous mode incurs significant error propagation in low SNR regime. 

The rest of this letter is organized as follows. System model of HARQ-CC-SCMA is introduced in Section \ref{Sys_model}. In Section \ref{AFGandLLR}, multi-user detection algorithms are proposed for HARQ-CC-SCMA. Simulation results are presented in Section \ref{Simuluation}. Finally, Section \ref{conclusion} concludes this letter.
\section{System Model}\label{Sys_model}
\begin{figure*}[htbp] 
    \centering 
    \includegraphics[width=0.90\textwidth]{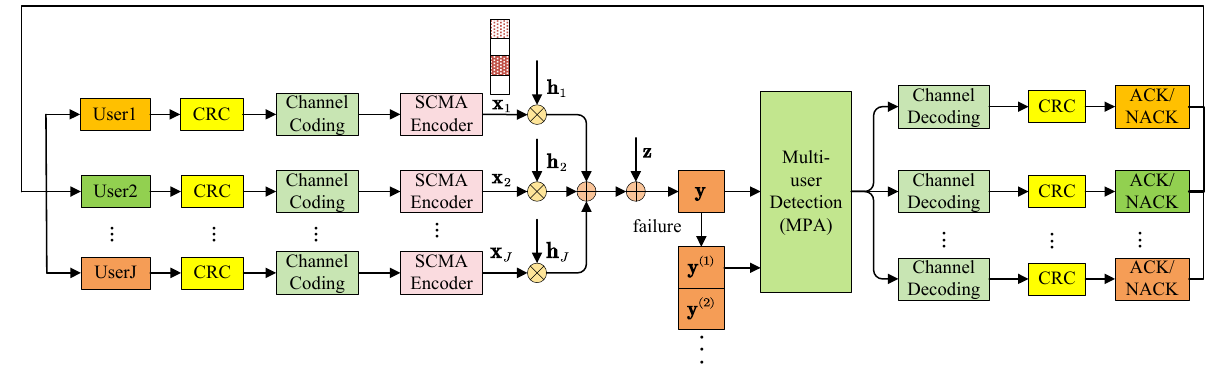} 
    \caption{The system model of HARQ-CC-SCMA.} 
    \label{Model} 
\end{figure*}

The system model of the proposed HARQ-CC-SCMA scheme is illustrated in Fig. \ref{Model}. We assume there are $J$ SCMA users. For each user $j$, cyclic redundancy check (CRC) bits are appended to the data bits of user $j$ and redundant bits are then inserted through channel encoder. Subsequently, the SCMA encoder maps the encoded bits of user $j$ to a $K$-dimensional sparse codeword $\mathbf{x}_j = (x_{1j},x_{2j},...,x_{Kj})^T$, which is chosen from a dedicated codebook of size $M$, where $(\cdot)^{T}$ is the transpose operator. As a consequence, the codewords of $J$ users are superimposed over $K$ orthogonal resource blocks. Due to the sparse nature of SCMA codebooks, each resource block is overlapped by only $d_f$ out of $J$ users and the user data is conveyed over $d_v<K$ resource blocks.

After receiving the superimposed signal corrupted by noise, multi-user detection is performed to calculate the LLR of each bit. The LLR is then used as the input of the channel decoder. The decoded data bits of each user are checked by performing CRC. The feedback of positive and negative acknowledgement (ACK/NACK) is sent to each user according to whether the CRC is passed or not. Once an NACK message is requested, user $j$ retransmits the last codeword $\mathbf{x}_j$. On the contrary, if an ACK message is fed back to user $j$, there are two transmission modes, i.e., synchronous and asynchronous, that determine the delivery of the next new message or the repetitive transmission of $\mathbf{x}_j$.



\subsubsection{Synchronous Transmission Mode}
In synchronous transmission mode, if any user receives an NACK feedback, all the users are needed to resend the same codeword until an ACK is received or the maximum number of transmission is reached. 
In the $q$-th HARQ round, the received signal at the base station (BS) is expressed as
\begin{equation}\label{Syn_y}
    \mathbf{y}^{(q)} = \sum\nolimits_{j=1}^{J} {{\rm diag}}(\mathbf{h}_{j}^{(q)})\mathbf{x} _{j}+\mathbf{z}_q,\, q\in[1,Q],
\end{equation}
where $ \mathbf{y}^{(q)} = (y_{1}^{ q},\cdots,y_{K}^{ q})^{T}$, $Q$ represents the maximum allowable number of transmissions, $\mathbf{h}_j^{(q)}=(h_{1j}^{q},h_{2j}^{q},...,h_{Kj}^{q})^T$ is the channel coefficient between user $j$ and BS, $\mathbf{z}_q$ is the additive white Gaussian noise (AWGN) with variance $N_0$, ${\rm diag}(\cdot)$ denotes the diagonal operation. 

\subsubsection{Asynchronous Transmission Mode}
In asynchronous transmission mode, once users receive ACK messages, they are allowed to deliver new data bits. However, the other users have to retransmit the failed codewords. In this case, each user has its own transmission index of HARQ rounds. We denote by $q_j$ the index of HARQ rounds of user $j$. Since different users may have different indices of HARQ rounds (i.e., $ q_1,\cdots,q_J\in[1,Q]$), \eqref{Syn_y} cannot be directly applied to define the signal model in asynchronous transmission mode. In the circumstance, the received signal at time slot $t$ is defined as
\begin{equation}
    \label{Asyn_y}
   \mathbf{y}^{(t)} =  \sum\nolimits_{j=1}^{J} {{\rm diag}}(\mathbf{h}_{j}^{(t)})\mathbf{x}^{(t)} _{j}+\mathbf{z}_{ t},
\end{equation}
where $\mathbf{h}_{j}^{(t)}$, $\mathbf{x}^{(t)} _{j}$, and $\mathbf{z}_{ t}$ denote the channel coefficient, the encoded symbol, and AWGN, respectively.



\section{Proposed HARQ-CC-SCMA Scheme}\label{AFGandLLR}
In this section, a novel HARQ-CC-SCMA scheme is proposed by considering two transmission modes (i.e., synchronous and asynchronous) and two multi-user detection strategies (i.e., FGA and LLRC). The implementation of the proposed HARQ-CC-SCMA scheme is detailed as follows.
\subsection{Synchronous Transmission Mode}
In synchronous transmission mode, new message for any user can be delivered if and only if all the users successfully recover their messages or the maximum number of transmissions is arrived. Therefore, the same superimposed packet is sent in all HARQ rounds. In what follows, FGA and LLRC are proposed for multi-user detection.
\subsubsection{FGA}\label{sec:fga_syn}
By considering multiple HARQ rounds and users, a large-scale factor graph is constructed to capture the connections between time-frequency resource blocks (referred to as factor nodes, i.e., FNs) and users (referred to as variable nodes, i.e., VNs), as shown in Fig. \ref{syn_AFG}. This amounts to aggregate multiple factor graphs together, each corresponding to one HARQ transmission. According to the mechanism of HARQ-CC-SCMA, regarding the aggregated factor graph after $q$ HARQ rounds in Fig. \ref{syn_AFG}, each VN is connected to $q\times d_v$ VNs and each FN is connected to $d_f$ VNs. By capitalizing on the sparsity of SCMA codebooks, MPA is leveraged to implement multi-user detection. Belief messages are passing between FNs and VNs to compute marginal probability distribution for each codeword. The process of message propagation on such a large-scale factor graph is elaborated as follows:
\begin{figure*}[htbp]
    \centering 
    \includegraphics[width=0.90\textwidth]{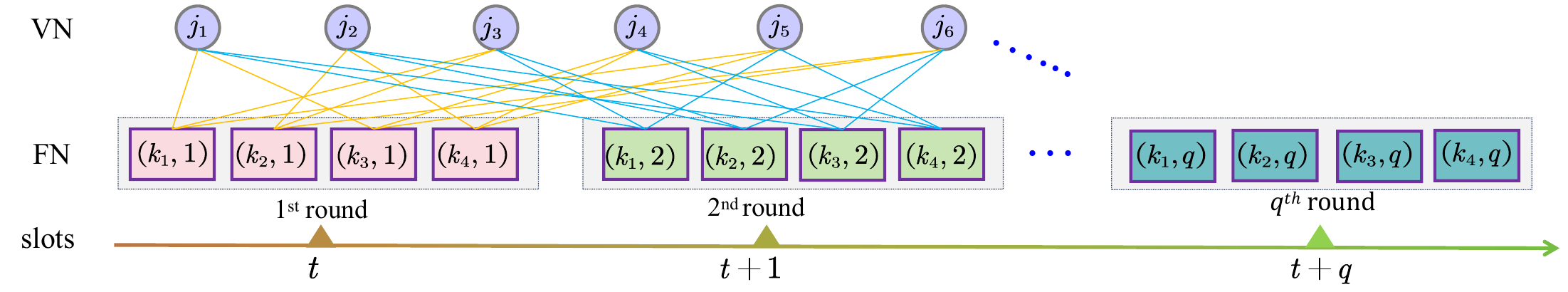} 
    \caption{An example of factor graph for modeling HARQ-CC-SCMA in synchronous transmission mode.} 
    \label{syn_AFG} 
\end{figure*}

\begin{enumerate}
    \item 
      In order to calculate the belief message passing from FN $(k,\mathfrak q)$ to VN $j$, FN $(k,\mathfrak q)$ needs to retrieve extrinsic information from the other $d_f-1$ neighbouring VNs. By multiplying the product of incoming belief messages by local likelihood function $\psi_{k,\mathfrak q}$ of FN $(k,\mathfrak q)$ and then marginalizing out VN $j$, the belief message passing from FN $(k,\mathfrak q)$ to VN $j$ at the $t$-th iteration precisely reads as
\begin{align}\label{FN-to-VN}
    m_{(k,\mathfrak q) \rightarrow j}^{(t)}(\mathbf{x}_{j})
    =&\sum\nolimits_{\left\{\mathbf{x}_{i}| i \in \mathcal N(k) \backslash j\right\}}\psi_{k,\mathfrak q}
    \notag\\
    &\times \prod\nolimits_{i \in \mathcal N(k) \backslash j} m_{i \rightarrow (k,\mathfrak q)}^{(t-1)}(\mathbf{x}_{i}),
\end{align}
where $\mathcal N(k)$ denotes the set of users occupying on the resource node $k$, $m_{i\rightarrow (k,\mathfrak {q})}^{(t-1)}(\mathbf{x}_{i})$ refers to the belief message passing from VN $i$ to FN $(k,\mathfrak q)$ at the $(t-1)$-th iteration, the notation $\mathcal N(k)\backslash j$ stands for all the VNs in $\mathcal N(k)$ except for VN $j$, and $\psi_{k,\mathfrak q}$ is the local likelihood function of FN $(k,\mathfrak q)$ that is given by
\begin{equation}\label{likelihood fun}
\begin{split}
    \psi_{k,\mathfrak q}&=p(y_{k}^{\mathfrak q}|\mathbf x_{j},j\in \mathcal N(k),N_{0})\\&\propto \exp \left(-\frac{1}{N_{0}}\left|y_{k}^{\mathfrak q}-\sum\nolimits_{j \in \mathcal N(k)}h_{kj}^{\mathfrak {q}}x_{kj}\right|^{2}\right).      
\end{split}
\end{equation}  

    \item  Unlike the traditional SCMA, the neighbours of each VN in the graph model of HARQ-CC-SCMA are not only associated with the current HARQ round but also the previous HARQ rounds. 
The belief message transmitted from VN $j$ to FN $(k,\mathfrak {q})$ at the $t$-th iteration is expressed as
\begin{equation}
    \label{VN-to-FN}
    m_{j\rightarrow (k,\mathfrak q)}^{(t)}(\mathbf{x}_{j})=\prod_{(i,\mathfrak q) \in \mathcal N(j) \backslash (k,\mathfrak q)}m_{(i,\mathfrak q) \rightarrow j}^{(t-1)}(\mathbf{x}_{j}),
\end{equation}
where $\mathcal N(j)$ denotes the time-frequency resource blocks overlapped by user $j$ in $q$ HARQ rounds.
\end{enumerate}


After $T$ iterations, the marginal probability distribution of codeword ${\bf x}_j$ is represented as
\begin{equation}
    \label{final_belief}
    p(\mathbf{x}_{j}) \propto  \prod\nolimits_{(k,\mathfrak {q}) \in \mathcal N(j) }m_{(k,\mathfrak {q}) \rightarrow j}^{(T)}(\mathbf{x}_{j}).
\end{equation}
With the estimated probability of the codeword at each user, the LLR for the $l$-th bit of the $j$-th user is calculated as
\begin{equation}
    \begin{split}
    \label{LLR}
    {\rm LLR}(b_{j}^{l})&=\log\frac{\Pr(b_{j}^{l}=1|\mathbf{y}^{(1)},\mathbf{y}^{(2)},\cdots,\mathbf{y}^{(q)})}{\Pr(b_{j}^{l}=0|\mathbf{y}^{(1)},\mathbf{y}^{(2)},\cdots,\mathbf{y}^{(q)})}\\    
                  &=  \log\frac{\sum_{\mathbf{x}_{j}\in \{{\rm{CB}}_{j}:b_{j}^{l}=1\}}p(\mathbf{x}_{j})}{\sum_{\mathbf{x}_{j}\in \{{\rm{CB}}_{j}:b_{j}^{l}=0\}}p(\mathbf{x}_{j})},   
    \end{split}              
\end{equation}
where $l\in\{1,\dots,\log_{2}M\}$, ${\rm{CB}}_{j}$ refers to the codebook of user $j$, and $\{{\rm{CB}}_{j}:b_{j}^{l}=a\}$ represents all the codewords in ${\rm{CB}}_{j}$ that are mapped from coded bit $b_{j}^{l}=a$. 
Afterwards, the generated LLRs are used as the input of channel decoder.
\renewcommand{\algorithmicrequire}{\textbf{Input:}} 
\renewcommand{\algorithmicensure}{\textbf{Output:}} 


\subsubsection{LLRC}
The complexity of FGA-based multi-user detection depends on the size of the factor graph, which obviously increases with the index of the current HARQ round. To get rid of this shortcoming, we resort to another alternative detection method that is devised by using the definition of LLR. Particularly, LLR represents the logarithm of the ratio of probabilities of $b_{j}^{l}=1$ being sent versus $b_{j}^{l}=0$ being sent. Accordingly, after $q$ HARQ rounds, it follows that
\begin{align}
    \label{LLR-combination}
    {\rm LLR}(b_{j}^{l})&=\log\left(\frac{\Pr(b_{j}^{l}=1|\mathbf{y}^{(1)},\mathbf{y}^{(2)},\cdots,\mathbf{y}^{(q)})}{\Pr(b_{j}^{l}=0|\mathbf{y}^{(1)},\mathbf{y}^{(2)},\cdots,\mathbf{y}^{(q)})}\right) \notag\\
    &= \log\left(\frac{\Pr(\mathbf{y}^{(1)},\mathbf{y}^{(2)},\cdots,\mathbf{y}^{(q)}|b_{j}^{l}=1)\Pr(b_{j}^{l}=1)}{\Pr(\mathbf{y}^{(1)},\mathbf{y}^{(2)},\cdots,\mathbf{y}^{(q)}|b_{j}^{l}=0)\Pr(b_{j}^{l}=0)}\right) \notag\\
    &=\sum\nolimits_{\mathfrak{q}=1}^{q} \log\left(\frac{\Pr(\mathbf{y}^{(\mathfrak{q})}|b_{j}^{l}=1)}{\Pr(\mathbf{y}^{(\mathfrak{q})}|b_{j}^{l}=0)}\right)\notag\\
    &=\sum\nolimits_{\mathfrak{q}=1}^{q} \log\left(\frac{\Pr(b_{j}^{l}=1|\mathbf{y}^{(\mathfrak{q})})}{\Pr(b_{j}^{l}=0|\mathbf{y}^{(\mathfrak{q})})}\right),                 
\end{align}
where the second and the last steps holds by using Bayesian formula, and the third step holds by assuming the independence of the received signals across different HARQ rounds and equal probability of $b_{j}^{l}$ being 0 and 1. With the aid of \eqref{LLR-combination}, LLRC is proposed to add up LLRs of erroneously received packets in previous HARQ rounds together with currently received packets, which is then used as the input of channel decoder. Clearly, it is of necessity to preserve the LLRs for failed users in all HARQ rounds until all the users successfully reconstruct their messages or the maximum number of transmissions is reached.

\subsection{Asynchronous Transmission Mode}
\begin{figure*}[htbp] 
    \centering 
    \includegraphics[height=0.15\linewidth]{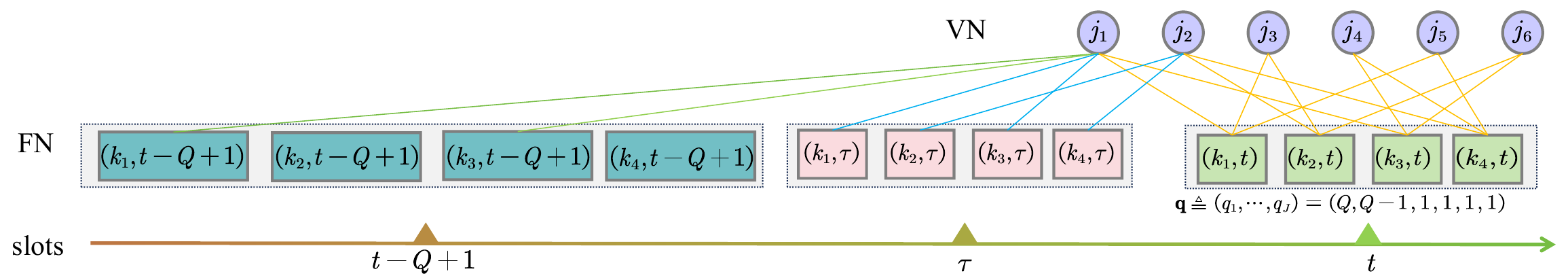} 
    \caption{An example of factor graph for modeling HARQ-CC-SCMA in asynchronous transmission mode.} 
    \label{FGAQ=3} 
\end{figure*}
In asynchronous transmission mode, there is no need to wait for all users to successfully decode their messages or for the maximum number of transmissions to reach before each user transmitting new data bits. In particular, upon receiving an ACK feedback from the BS or arriving the maximum number of HARQ rounds, user $j$ initiates the delivery of new data bits in the subsequent time slot. Otherwise, user $j$ keeps retransmitting the same message in the next HARQ round. 
Therefore, different superimposed signals might be retransmitted. In order to adapt to asynchronous transmission mode, the FGA and LLRC are redesigned for multi-user detection in this case.
\subsubsection{FGA}\label{Asyn.-FGA}
At time slot $t$, the indices of HARQ rounds for $J$ users are defined as ${\bf q} = (q_1,\cdots,q_J)^T$. We denote by $\mathcal C$ the set of the packets transmitted by $J$ users at the current time slot $t$. In order to recover $\mathcal C$, it is imperative to exploit all the previously received signals relevant to $\mathcal C$ for joint detection. Accordingly, ${{\bf{y}}^{( t-\bar q+1)}},\cdots, {{\bf{y}}^{( t)}}$ are employed to construct a large size factor graph, where $\bar q = \max\{q_1,\cdots,q_J\}$. Herein, it is worth mentioning that ${{\bf{x}}_j^{(t-q_j+1 )}}=\cdots={{\bf{x}}_j^{(t )}}$ according to the definition of ${\bf q}$. Moreover, ${{\bf{y}}^{(\tau )}}$ for $\tau \in[t-\bar q+1,t]$ might contain the previous packets associated with old data bits that have been successfully decoded or dropped. For notational convenience, we define $\mathcal S$ and $\mathcal D$ as the sets of the packets that have been recovered and discarded before time slot $t$, respectively. The presence of discarded packets $\mathcal D$ are due to the fact that these packets cannot be reconstructed within $Q$ transmissions. Accordingly, the received signal ${{\bf{y}}^{(\tau )}}$ can be rewritten as
\begin{align}
    \label{y_fail}
{{\bf{y}}^{(\tau )}} =& \underbrace {\sum\nolimits_{{\bf{x}}_j^{(\tau )} \in \mathcal C} {{\rm{diag}}({\bf{h}}_j^{(\tau)}){\bf{x}}_j^{(\tau )}} }_{\rm{desired~signal}} + \underbrace {\sum\nolimits_{{\bf{x}}_j^{(\tau )} \in \mathcal S} {{\rm{diag}}({\bf{h}}_j^{(\tau)}){\bf{x}}_j^{(\tau )}} }_{\rm{recovered~signal}}\notag\\
& + \underbrace {\sum\nolimits_{{\bf{x}}_j^{(\tau )} \in \mathcal D} {{\rm{diag}}({\bf{h}}_j^{(\tau )}){\bf{x}}_j^{(\tau )}} }_{\rm{interfering~signal}} + {{\bf{z}}_\tau }.
\end{align}
By invoking MPA, the FGA algorithm introduced in Section \ref{sec:fga_syn} can be similarly applied to asynchronous transmission mode. The only difference essentially originates from local likelihood function of FN $(k,\tau)$. More specifically, the term of recovered signal in \eqref{y_fail} can be cancelled out and the term of interfering signal should be treated as noise. We thus define the equivalent noise term as $\tilde {\bf z}_\tau=\sum\nolimits_{{\bf{x}}_j^{(\tau )} \in \mathcal D} {{\rm{diag}}({\bf{h}}_j^{(\tau )}){\bf{x}}_j^{(\tau )}} + {{\bf{z}}_\tau }$, where the variance matrix $\bf C$ of $\tilde {\bf z}_\tau$ is expressed as
\begin{align}
    \label{interface}
           &  {\bf{C}} = \mathbb E\left\{ {{{\widetilde {\bf{z}}}_\tau }{{\widetilde {\bf{z}}}_\tau }^H} \right\}\notag\\
            %
& = {N_0}{\bf{I}} + \sum\nolimits_{{\bf{x}}_j^{(\tau )} \in {\cal D}} {\mathbb E\left\{ {{\rm{diag}}({\bf{x}}_j^{(\tau )}){{\left( {{\rm{diag}}({\bf{x}}_j^{(\tau )})} \right)}^H}} \right\}}\notag\\
& = {N_0}{\bf{I}} + \sum\nolimits_{{\bf{x}}_j^{(\tau )} \in {\cal D}} {\frac{1}{M}\sum\limits_{{{\bf{x}}_j} \in {\rm{C}}{{\rm{B}}_j}} {{\rm{diag}}\left( {{{\left| {{x_{1j}}} \right|}^2}, \cdots ,{{\left| {{x_{Kj}}} \right|}^2}} \right)} }\notag\\
&\triangleq  {\rm{diag}}(\sigma_1^2,\cdots,\sigma_K^2),
\end{align}
where the second step holds by using independent fading channels with unity average power. The corresponding local likelihood function of FN $(k,\tau)$ is thus given by
\begin{equation}\label{likelihood_funasy}
    \psi_{k,\tau}
    =\frac{1}{{\pi \sigma _k^2}}\exp \left( { - \sigma _k^{ - 2}{{\left| {\tilde y_k^\tau  - \sum\limits_{{\bf{x}}_j^{(\tau )} \in {\cal C} \wedge j \in \mathcal N(k)} {h_{kj}^\tau x_{kj}^\tau } } \right|}^2}} \right),
\end{equation}  
where $\tilde y_k^\tau$ is obtained by subtracting  the recovered signal from ${{\bf{y}}^{(\tau )}} $. 
The propagation of belief message is the same as the FGA in synchronous transmission mode. As shown in Fig. \ref{FGAQ=3}, $\mathbf{q}=(Q,Q-1,1,1,1,1)$ is taken as an example, users 1 and 2 are engaging into retransmissions, the messages of users 4, 5, and 6 were successfully recovered at time slot $t-1$, and the message of user 3 is discarded at time slot $t-1$. From the foregoing analysis, with regard to the factor graph of asynchronous HARQ-CC-SCMA, some edges between the time-frequency resource blocks and successful/discarded packets are removed. This apparently differs from the factor graph built in synchronous mode.


\subsubsection{LLRC}
Unlike LLRC in synchronous mode, LLRC in asynchronous mode adds up LLRs in the last $q_j$ HARQ rounds for coded bit $b_j^l$.

\section{Simulation Results}\label{Simuluation}
The simulation results are presented in this section. For comparison, the traditional Type-I HARQ and HARQ-IR assisted SCMA schemes are used as baselines. For illustration, the simulation parameters are set as $J = 6$, $K = 4$, $M = 4$, and $Q=2$ unless otherwise stated. We assume that there are 10000 HARQ rounds in total and 200 information bits in each round is delivered by each user. Moreover, the codebooks of SCMA in Table III of \cite{9145202} are used in the simulation. Turbo code with rate ${1}/{2}$ and 16 bits CRC are used for channel coding. In addition, the algorithm of MAX-Log-MPA in \cite{8967122} is adopted for SCMA detection and the maximum number of iterations of MAX-Log-MPA is set as $6$. 

\begin{figure*}
\begin{minipage}[b]{.29\linewidth}
    \centering
    \includegraphics[width=0.9\textwidth]{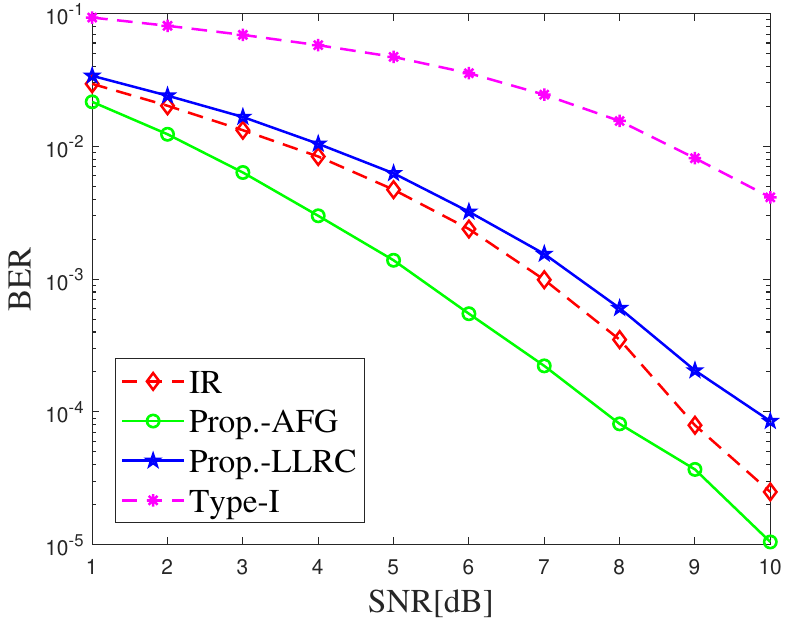}
    \caption{BER of HARQ-assited SCMA schemes in synchronous mode.}
    \label{Syn-sim}
\end{minipage} 
\hfill
\begin{minipage}[b]{.29\linewidth}
    \centering
    \includegraphics[width=0.9\textwidth]{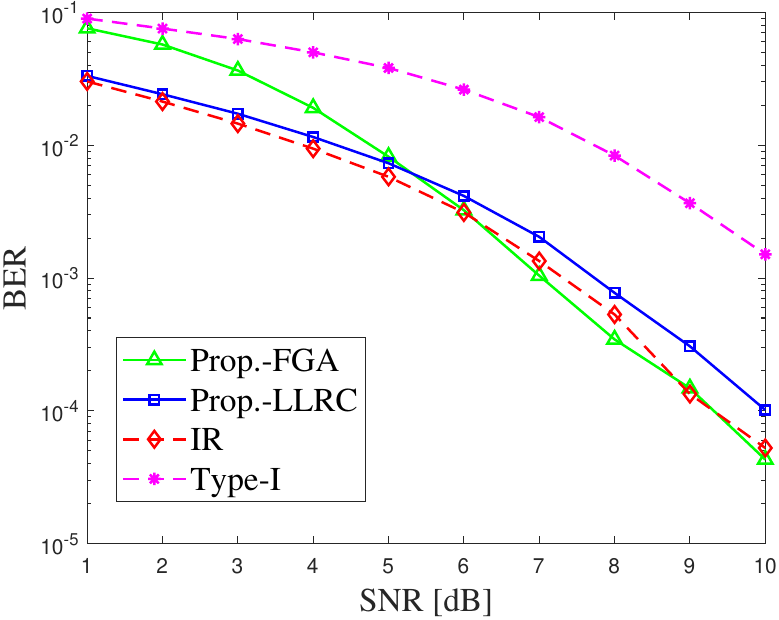} 
    \caption{BER of HARQ-assisted SCMA schemes in asynchronous mode.}
    \label{Asyn-sim}
\end{minipage} 
\hfill
\begin{minipage}[b]{.29\linewidth}
    \centering
    \includegraphics[width=0.9\textwidth]{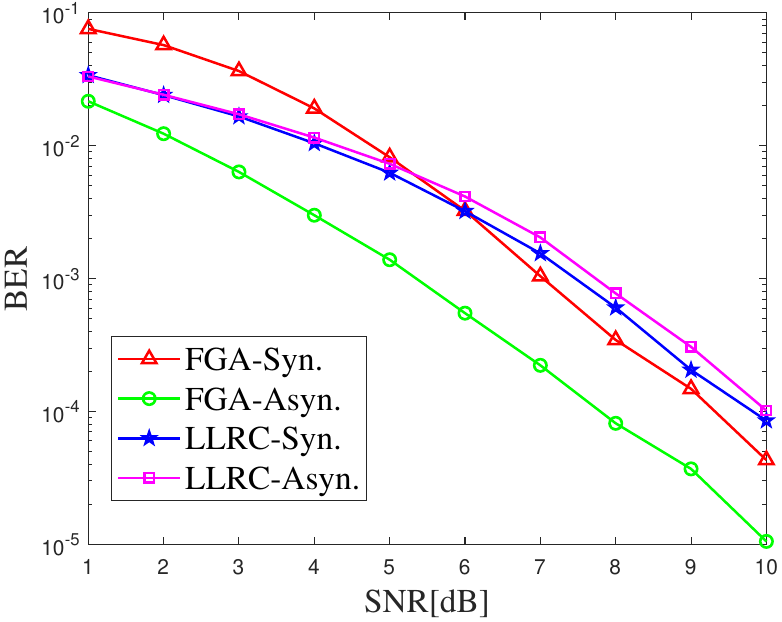}
    \caption{Comparison between synchronous and asynchronous modes.}
    \label{compare}
\end{minipage} 
\hfill
\end{figure*}

Fig. \ref{Syn-sim} investigates bit error rate (BER) of HARQ-SCMA schemes in synchronous transmission mode. It can be observed from Fig. \ref{Syn-sim} that the proposed HARQ-CC-assisted SCMA with FGA (labeled as Prop.-FGA) has the lowest BER. This is achieved at the expense of high multi-user detection complexity, which increases with the scale of factor graph. Moreover, it can be seen that the Prop.-FGA performs better than HARQ-IR-SCMA (labeled as IR). This indicates that the joint SCMA detection between HARQ rounds prior to channel decoding is more effecient than the joint channel decoding between HARQ rounds after SCMA detection. Besides, the proposed HARQ-CC-assisted SCMA with LLRC (labeled as Prop.-LLRC) is inferior to HARQ-IR-SCMA. This is because Prop.-LLRC is essentially a simple combination of all the LLRs from the current and previous HARQ rounds, which results in a low complexity. In addition, it is shown in Fig. \ref{Syn-sim} that Type-I HARQ-assisted SCMA (labeled as Type-I) performs the worst. This is due to the fact that only the signal received in the current HARQ round is used for decoding and the useful information contained in erroneously received packets is wasted.

Fig. \ref{Asyn-sim} plots BER of HARQ-SCMA schemes versus SNR in synchronous transmission mode. The simulation results reveal that the Prop.-FGA performs worse than Prop.-LLRC at low SNR. This is because there is a large number of decoding errors at low SNR. Failed packets in the superimposed signal cannot be cancelled out, which yields a large aggregated user interference, consequently resulting in severe error propagation. Whereas, the error propagation can be relieved  in high SNR owing to the occurrence of less decoding errors. It can be seen that the Prop.-FGA almost reaches a comparable BER performance to HARQ-IR-SCMA. This demonstrates that the Prop.-FGA is more susceptible to the interference from failed users by comparing to other algorithms. In addition, the relationship among Prop.-LLRC, IR, and Type-I in Fig. \ref{Asyn-sim} is similar to that observed in Fig. \ref{Syn-sim}.


Fig. \ref{compare} shows a BER performance comparison between HARQ-CC-SCMA schemes in synchronous and asynchronous transmission modes. It is found that FGA in synchronous mode (labeled as FGA-Asyn.) performs much better than FGA in asynchronous mode (labeled as FGA-Syn.) in terms of BER. This is owing to the fact that asynchronous transmission mode yields serious error propagation from failed users. However, with regard to LLRC, LLRC in synchronous mode (labeled as LLRC-Syn.) performs a little bit better than FGA in asynchronous mode (labeled as LLRC-Asyn.). This reveals that the LLRC-based algorithm is insensitive to the transmission mode and exhibits a better stability by comparing to FGA-based one.

\section{Conclusion}\label{conclusion}
This letter proposed a novel HARQ-CC-SCMA scheme, which takes into account both synchronous and asynchronous transmission modes. In addition, FGA and LLRC have been developed to exploit the useful information of erroneously received codewords for multi-user detection. Finally, simulation results have proved that FGA outperforms both LLRC and HARQ-IR-SCMA in synchronous mode. However, LLRC performs better than FGA at low SNR in asynchronous mode. This is due to the severe error propagation in the circumstance.


\bibliographystyle{IEEEtran}
\bibliography{reference}

\begin{thebibliography}{10}
\providecommand{\url}[1]{#1}
\csname url@samestyle\endcsname
\providecommand{\newblock}{\relax}
\providecommand{\bibinfo}[2]{#2}
\providecommand{\BIBentrySTDinterwordspacing}{\spaceskip=0pt\relax}
\providecommand{\BIBentryALTinterwordstretchfactor}{4}
\providecommand{\BIBentryALTinterwordspacing}{\spaceskip=\fontdimen2\font plus
\BIBentryALTinterwordstretchfactor\fontdimen3\font minus \fontdimen4\font\relax}
\providecommand{\BIBforeignlanguage}[2]{{%
\expandafter\ifx\csname l@#1\endcsname\relax
\typeout{** WARNING: IEEEtran.bst: No hyphenation pattern has been}%
\typeout{** loaded for the language `#1'. Using the pattern for}%
\typeout{** the default language instead.}%
\else
\language=\csname l@#1\endcsname
\fi
#2}}
\providecommand{\BIBdecl}{\relax}
\BIBdecl

\bibitem{10041914}
M.~Chafii, L.~Bariah, S.~Muhaidat, and M.~Debbah, ``Twelve scientific challenges for 6{G}: Rethinking the foundations of communications theory,'' \emph{IEEE Commun. Surveys \& Tuts.}, vol.~25, no.~2, pp. 868--904, Feb. 2023.

\bibitem{9409837}
L.~Yu, Z.~Liu, M.~Wen, D.~Cai, S.~Dang, Y.~Wang, and P.~Xiao, ``Sparse code multiple access for 6{G} wireless communication networks: Recent advances and future directions,'' \emph{IEEE Commun. Stand. Mag.}, vol.~5, no.~2, pp. 92--99, Jun. 2021.

\bibitem{9782313}
S.~Chaturvedi, Z.~Liu, V.~A. Bohara, A.~Srivastava, and P.~Xiao, ``A tutorial on decoding techniques of sparse code multiple access,'' \emph{IEEE Access}, vol.~10, pp. 58\,503--58\,524, Jun. 2022.

\bibitem{7500115}
Y.~Long, Z.~Chen, Z.~Guo, and J.~Fang, ``A novel {HARQ} scheme for {SCMA} systems,'' \emph{IEEE Wireless Commun. Lett.}, vol.~5, no.~5, pp. 452--455, Oct. 2016.

\bibitem{8322726}
J.~Lian, S.~Zhou, X.~Zhang, and Y.~Wang, ``An improved {HARQ} scheme for {SCMA} under random access,'' in \emph{Proc. IEEE Int. Conf. Comput. Commun. (ICCC)}, Chengdu, China, Dec. 2017, pp. 1163--1167.

\bibitem{10295166}
K.~Lai, Z.~Liu, J.~Lei, G.~Chen, P.~Xiao, and L.~Wen, ``Sparse code multiple access with enhanced {K}-repetition scheme: Analysis and {D}esign,'' \emph{IEEE Trans. Wireless Commun.}, vol.~99, no.~99, pp. 1--1, Oct. 2023.

\bibitem{zhu2018rateless}
M.~Zhu, Q.~He, R.~Zhang, and B.~Bai, ``Rateless coding based incremental redundancy {HARQ} scheme for {SCMA} systems,'' \emph{Mob. Netw. Appl.}, vol.~23, pp. 1028--1034, Aug. 2018.

\bibitem{e25060930}
M.~Guan, M.~Zhu, and B.~Bai, ``Windowed joint detection and decoding with {IR-HARQ} for asynchronous {SCMA} systems,'' \emph{Entropy}, vol.~25, no.~6, pp. 1--13, 2023.

\bibitem{9745558}
H.~Zhang, Z.~Liao, Z.~Shi, G.~Yang, Q.~Dou, and S.~Ma, ``Performance analysis of {MIMO-HARQ} assisted {V2V} communications with keyhole effect,'' \emph{IEEE Trans. Commun.}, vol.~70, no.~5, pp. 3034--3046, May 2022.

\bibitem{9145202}
K.~Deka, M.~Priyadarsini, S.~Sharma, and B.~Beferull-Lozano, ``Design of {SCMA} codebooks using differential evolution,'' in \emph{Proc. IEEE Int. Conf. Commun. Workshops (ICC Workshops)}, Dublin, Ireland, Jun. 2020, pp. 1--7.

\bibitem{8967122}
C.~Zhang, C.~Yang, X.~Pang, W.~Song, W.~Xu, S.~Zhang, Z.~Zhang, and X.~You, ``Efficient sparse code multiple access decoder based on deterministic message passing algorithm,'' \emph{IEEE Trans. Veh. Technol.}, vol.~69, no.~4, pp. 3562--3574, Apr. 2020.

\end{thebibliography}

\end{document}